\documentclass[print]{aa}

\usepackage{graphicx}

\usepackage{natbib}

\begin{document}

\title{On the missing $2175$~\AA{}-bump in the Calzetti extinction curve}


\author{J\"org Fischera \and Michael Dopita}
\institute{
	Research School of Astronomy \& Astrophysics, Mount Stromlo Observatory, Cotter Road, 
	Weston Creek, ACT 2611, Australia}


\abstract{}
	{The aim of the paper is to give a physical explanation of the absence of the feature 
	in the Calzetti extinction curve.}
	{We analyze the dust attenuation of a homogeneous source seen through a distant inhomogeneous
	distant screen. The inhomogeneities are described through an idealized isothermal turbulent medium
	where the probability distribution function (PDF) of the column density is log-normal. In addition 
	it is assumed that  below a certain critical column density
	the carriers of the extinction bump at 2175~\AA{} are being destroyed
	by the ambient UV radiation field.}
	{Turbulence is found to be a natural explanation not only of the flatter curvature of the Calzetti
	extinction curve but also of the missing bump provided the critical column density is
	$N_{\rm H}\ge 10^{21}~{\rm cm^{-2}}$. The density contrast needed to explain both
	characteristics is well consistent with the Mach number of the cold neutral medium of our own Galaxy
	which suggests a density contrast $\sigma_{\rho/\left<\rho\right>}\approx 6$.}


\maketitle

\section{Introduction}

The ability of interstellar dust grains to attenuate light through scattering and absorption can lead to large uncertainties
in the determination of the crucial physical parameters of galaxies. These parameters not only help determine the
evolution of the galactic system, but are also important for the determination of the star formation rate as
function of red-shift, a key parameter in understanding the evolution of
the universe as a whole. The dust correction is complicated by
several effects as the contribution of scattered light, the geometry, the mixture of the dust with the sources,
and the inhomogeneous structure of the interstellar medium all affect the global attenuation of the galactic starlight. 
These effects produce an attenuation curve
which may be quite different from a pure extinction curve derived for an individual star.

It is maybe not too surprising that the galactic reddening curve $E(\lambda-V)/E(B-V)$ derived for star-burst galaxies
\citep{Calzetti2001} shows two significant deviations from the extinction curve known for the diffuse interstellar
medium (ISM) of our own galaxy. First the reddening at long wavelengths is lower 
which points to a larger absolute-to-relative extinction $R_V$ and to flatter extinction curve $A_\lambda/A_V$ and second the 
2175~\AA{} feature which is very prominent in the extinction curves derived for the Milky Way or the Large Magellanic Cloud (LMC) seems to be rather weak or absent. The smoothness makes the Calzetti curve similar to the 
extinction curve derived for the bar of the Small Magellanic Cloud (SMC) \citep{Gordon1998}. However, the overall curvature in the optical is
steeper in the SMC with an $R_V$-value even slightly lower ($2.74\pm0.13$, \citet{Gordon2003}) in respect to the Milky Way (3.1, \citet{Fitzpatrick1999}). 

A flatter curvature of a pure foreground extinction points in general to a larger grain population as is 
inferred in the case of the Orion region, for example. A flattening of the effective extinction curve 
for extended sources can also be produced by the in-homogeneity of the dusty interstellar medium. 
The reason for this lies in the fact that a non-homogeneous medium
is less optically thick than a homogeneous distribution of matter 
and that the reduction in the effective extinction increases with optical thickness.
It has been found based on radiative transfer calculations that 
a clumpy shell can reproduce the Calzetti curve if dust properties are consistent with the smooth SMC bulk 
extinction curve \citep{Gordon1997,Witt2000}. 
We have shown \citep{Fischera2003} that the overall flatter curvature of the Calzetti curve can be naturally explained by the turbulent nature of the ISM (paper~I). The density contrast needed to produce the flattening is found
to be consistent with the velocity dispersion of the cold neutral medium (CNM) as indicated by 
CO observations.

The situation with regard to the absence of the peak at 2175~\AA{} is rather different. It is thought that the peak is 
caused by $\pi$-electron resonance produced in small carbonaceous particles which include graphenes,
polycyclic aromatic hydrocarbons (PAH) and possibly small graphite grains
\citep{Li2001,Weingartner2001,DraineLi2007,Fischera2008}. The UV light is thought to excite the skeleton 
vibration modes of the molecules which produce in case of PAH molecules the broad emission features
seen in the near infrared. 
The analysis of the diffuse galactic emission \citep{Witt1973,Lillie1976,Morgan1976} or reflection nebula \citep{Witt1982,Witt1992,Calzetti1995} suggests that the feature is predominantly or even completely caused by absorption.
If the observed light contains a considerable amount of scattered emission the peak strength in the effective
extinction curve would even increase. Geometrical effects for the absence of the feature in the Calzetti curve
can therefore be excluded as verified by detailed radiative transfer calculations \citep{Gordon1997,Witt2000}.

The only possible explanation for the smooth Calzetti curve seems to be the destruction 
of the carriers of the peak caused by the strong UV radiation field. This process has been discussed
in the context of star burst galaxies by \citet{Dopita2005}, and has been extensively modeled by several authors \citep{Omont1986,Allamandola1989,Leger1989,Allain1996a,Allain1996b,Page2003} 
A similar interpretation was given for the absence of the peak in the SMC bar extinction curve. The carriers are
thought to be destroyed by the pervasive UV radiation field caused by the lower dust content resulting from a ten times 
\citep{Russel1992} lower metallicity in the ISM of the SMC \citep{Gordon1998}. 
The destruction in the ISM  cannot be complete as one individual sight line shows a clear extinction bump \citep{Lequeux1982,Gordon1998,Gordon2003}.
A complete destruction of the carriers also seems to be in contradiction with the observed emission spectra
of starburst galaxies as they still show the prominent PAH emission features. 

This paper is number IV in a series of papers where we analyze the attenuation characteristics caused
by a dusty turbulent medium. In the first paper (paper I) we addressed the problem if a turbulent medium can reproduce
the flatter curvature of the Calzetti extinction curve. In the following paper \citep{Fischera2004a} we provided a model of the
isothermal turbulent screen and showed how the statistical properties (the 1-point and the 2-point statistic) 
of the column density are related to the statistical properties of the local densities (paper II). We have applied this model
\citep{Fischera2005a} to analyze in detail the attenuation caused by a distant turbulent screen (paper III).
In this current paper we investigate under which circumstances the 2175~\AA{} absorption feature can be suppressed, while
at the same time not removing all its carriers through PAH photo-dissociation.

\section{Model}

In any model, the geometry of the dust with respect to an extended source of photons is crucial in determining 
the received intensity at any wavelength. As in our previous paper~III we will apply the
geometry of a distant turbulent dusty screen.

\subsection{\label{sect_turb}The turbulent slab}

For the dusty screen we assume an isothermal turbulent medium.
Turbulence produces a broad distribution of the local density $\rho$ which is because of the dependence of the densities
from their neighboring density values described 
in the absence of gravity through a log-normal function. If we consider the normalized values $\hat x=x/\left<x\right>$ where $\left<x\right>$ is the mean the probability distribution function (PDF) is given by:
\begin{equation}
	\label{eq_lognormal}
 	p(\ln \hat x) = \frac{1}{\sqrt{2\pi}\sigma_{\ln \hat x}} \exp\left[-\frac{1}{2\sigma_{\ln \hat x}^2}\left(\ln \hat x-\ln \hat x_0\right)^2\right]
\end{equation}
with $\ln\hat x_0 = -\frac{1}{2}\sigma^2_{\ln \hat x}$ where $\sigma_{\ln\hat\rho}$ is the standard deviation of the log-normal
function. The standard deviation of the log-normal function is directly related to the standard deviation of the normalized values:
\begin{equation}
	\sigma_{\hat x} = \sqrt{e^{\sigma^2_{\ln \hat x}}-1}
\end{equation}
In case of the local density the density contrast is according to \citet{Padoan1997} directly related to the Mach number $\delta v/c M$ (where
$\delta v$ is the velocity dispersion and $c$ the sound speed) in the medium
with $\sigma_{\hat \rho} = \beta M$ with $\beta=1/2$. For the cold neutral medium (CNM) in our galaxy the CO measurements
for example imply a Mach number $M \approx 12$ which provides $\sigma_{\hat \rho}\approx 6$.

The log-normal function is very robust and becomes only skewed in in the presence of self-gravity.
This produces higher probabilities of encountering high densities since those values are located in the massive
clouds which are gravitationally more stable against the turbulent motion.

The log-normal function eq.~\ref{eq_lognormal} also approximately applies to the distribution of the normalized column density $\xi = N_{\rm H} /\left<N_{\rm H}\right> $ through an idealized turbulent medium \citet{Fischera2003}. We have shown how 
the ratio of the standard deviation
of the column density and the standard deviation of the local density depends apart from the thickness $\Delta$ of the turbulent slab which is conveniently measured in terms of the maximum turbulent scale $L_{\rm max}$ also on the structure of the local density.
In the turbulent medium the structure is described through a simple power law of the local density in Fourier space 
$P(\rho(k))=k^{n}$ where $k$ is the wavenumber. In the limit of a thin slab ($\Delta/L_{\rm max}\ll 1$) the standard deviations become equal. 
In the limit of a thick screen ($\Delta/L_{\rm max}\gg 1$) the variance of the column density is inversely proportional to the thickness.
Assuming a simple power law this limit is given by:
\begin{equation}
	\label{eq_sigma}
	\sigma_\xi = \sigma_{\rho/\left<\rho\right>} \sqrt{\frac{n+3}{2n+4}\frac{L_{\rm max}}{\Delta}}.
\end{equation}
where $n<-3$. The value $n=-11/3$ is consistent with Kolmgorov turbulence. As shown in paper~II
the asymptote also provides accurate results at $\Delta/L_{\rm max}=1$.

In this model the standard deviation of the column density is not clearly defined as the value depends not only on 
the turbulence of the medium (Mach number $M$) but also on the thickness of the slab $\Delta/L_{\rm max}$. The same
distribution function can be obtained through different assumptions of the turbulence and the slab thickness.
In this work we consider therefore the distribution function of the column density assuming different values of the total mean
column density and the standard deviation $\sigma_{\ln \xi}$.

\subsection{The effective extinction}

We assume that the carriers of the 2175~\AA{} bump are being destroyed below a certain column density $[N_{\rm H}]_{\rm crit}=\xi_{\rm crit}\left<N_{\rm N}\right>$ but intact at higher column densities.
The corresponding extinction coefficients in the medium below and above this
critical column density are distinguished as $\kappa_\lambda^{(1)}$ and $\kappa_\lambda^{(2)}$, respectively.

The effective or effective extinction of a homogeneous light source seen through a turbulent, or in general non-homogeneous, dusty screen is given by the mean of the extinction values
\begin{equation}
	\tau_\lambda^{\rm eff} = -\ln \left<e^{-\tau_\lambda}\right> 
\end{equation}
where $\tau_\lambda = \kappa_\lambda N_{\rm H}$ is the optical depth.
For the idealized turbulent screen the effective extinction is then given by:
\begin{eqnarray}
	\label{eq_taueff}
	e^{-\tau^{\rm eff}_\lambda} &=& \int\limits_{-\infty}^{y_{\rm crit}}{\rm d}y\,e^{-\left<\tau^{(1)}_\lambda\right>e^y}\,p(y)\nonumber\\
		&&+ e^{-\Delta\left<\tau_\lambda\right>e^{y_{\rm crit}}}\int_{y_{\rm crit}}^{\infty}{\rm d}y\,e^{-\left<\tau^{(2)}_\lambda\right>e^y}\,p(y) 
\end{eqnarray}
where $\left<\tau_\lambda^{(i)}\right>=\kappa_\lambda^{(i)}\left<N_{\rm H}\right>$ is the mean extinction 
and where we use the abbreviation $y = \ln \xi$ and $y_{\rm crit}=\ln \xi_{\rm crit}$. Further, we have difference
in optical depth $\Delta\left<\tau_\lambda\right>=\left<\tau_\lambda^{(1)}\right>-\left<\tau_\lambda^{(2)}\right>$.

\begin{figure}[htbp]
	\includegraphics[width=\hsize]{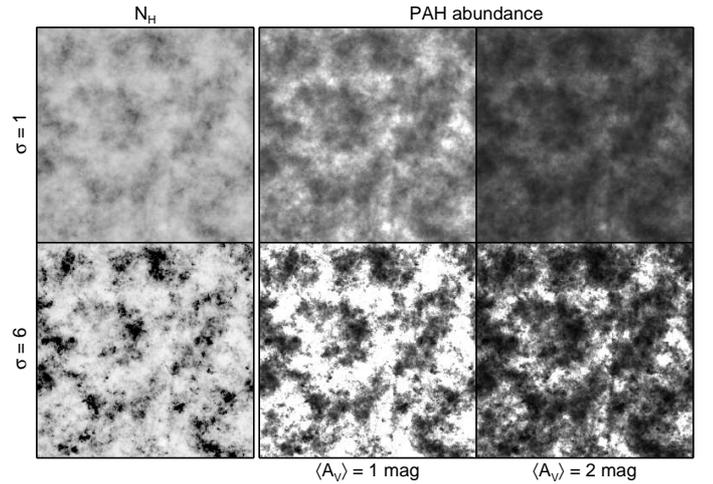}
	\caption{\label{fig_visu}
	Visualization of the effect caused by PAH destruction below a critical column density 
	$[N_{\rm H}]_{\rm crit}=10^{21}~{\rm cm^{-2}}$. The density structure is derived using a cube of $256^3$ Pixel, a power
	index of $n=-10/3$ and a maximum length scale of $0.4$ times the cube size. To the left the column density for two standard 
	deviations $\sigma_{\rho/\left<\rho\right>}$ of the local density is shown. The images to the right for each turbulent
	cube gives the PAH abundance along individual sight lines for two assumptions of the extinction $A_V$ through
	the cube. The white fields correspond to no PAH molecules while the black fields have the normal PAH abundance
	necessary to produce the extinction bump at 2175~\AA{}. 
	}
\end{figure}

The effect caused by the destruction of PAH molecules below a column density of $[N_{\rm H}]_{\rm crit}=10^{21}~{\rm cm^{-2}}$
is visualized in Fig.~\ref{fig_visu} for two different assumptions of the density contrast and the mean extinction 
$\left<A_V\right>$ through the 
slab. The density contrasts 1 and 6 correspond, assuming a power $n=-10/3$, to a standard deviation $\sigma_{\ln \xi}$ of the log-normal density distribution of the column density of approximately $0.22$ and $1.01$.

The turbulence produces a large variety of column densities. In high turbulent media the medium is
compressed to small volumes of high densities which appear in projection as individual clouds leaving large areas of low column densities. An appreciable fraction of the total projected area is completely free of the carriers of the 2175~\AA{} absorption feature.
As more light is transmitted in these regions, the size of the bump in the overall attenuation curve is reduced as well.

On the other hand the carrier abundance of the bump is higher in a turbulent medium as the dust is now 
found predominantly in dense clouds where the carriers become because of the high column densities save against destruction.

\subsection{Extinction curve}

For the calculations we adopt the extinction curve as provided by \citet{Fitzpatrick1999}. In the model the feature
at 2175~\AA{} is described by a Drude model which we subtracted to obtain the curve for a medium where the carriers
are destroyed. For simplicity we assume that the absolute-to-relative extinction does not depend 
on the carrier destruction.
To analyze the effect of the destruction on the attenuation curve we consider the 
extinction coefficients at peak frequency. The corresponding values without and with the absorption feature are given by:
\begin{eqnarray}
	\label{eq_extcoeffpeak}
	\kappa_{0.22}^{(1)}&=& 5.75 / (5.8\times 10^{21})~{\rm cm^{-2}},\nonumber\\
	\kappa_{0.22}^{(2)}&=& 8.78 / (5.8\times 10^{21})~{\rm cm^{-2}}.
\end{eqnarray}
As peak strength we define
\begin{equation}
	\frac{\Delta A_{0.22}}{E(B-V)}
\end{equation}
where $\Delta A_{0.22}$ is the difference of the extinction at peak frequency with and without absorption feature.
In case of the Fitzpatrick curve this value is $c_3/\gamma^2 = 3.30$ where $c_3=3.23$ is the bump strength and 
$\gamma=0.99$ the bump width \citep{Fitzpatrick1999}.

\section{Results}

First we show that a turbulent medium with the combination of additional destruction can indeed lead to a reduced
feature at 2175~\AA{}. 
We will then analyze more quantitative the requirements to produce smooth attenuation curves.
Approximations of the effective optical depths are given in App.~\ref{app}.

\begin{figure*}[htbp]
	\includegraphics[width=0.49\hsize]{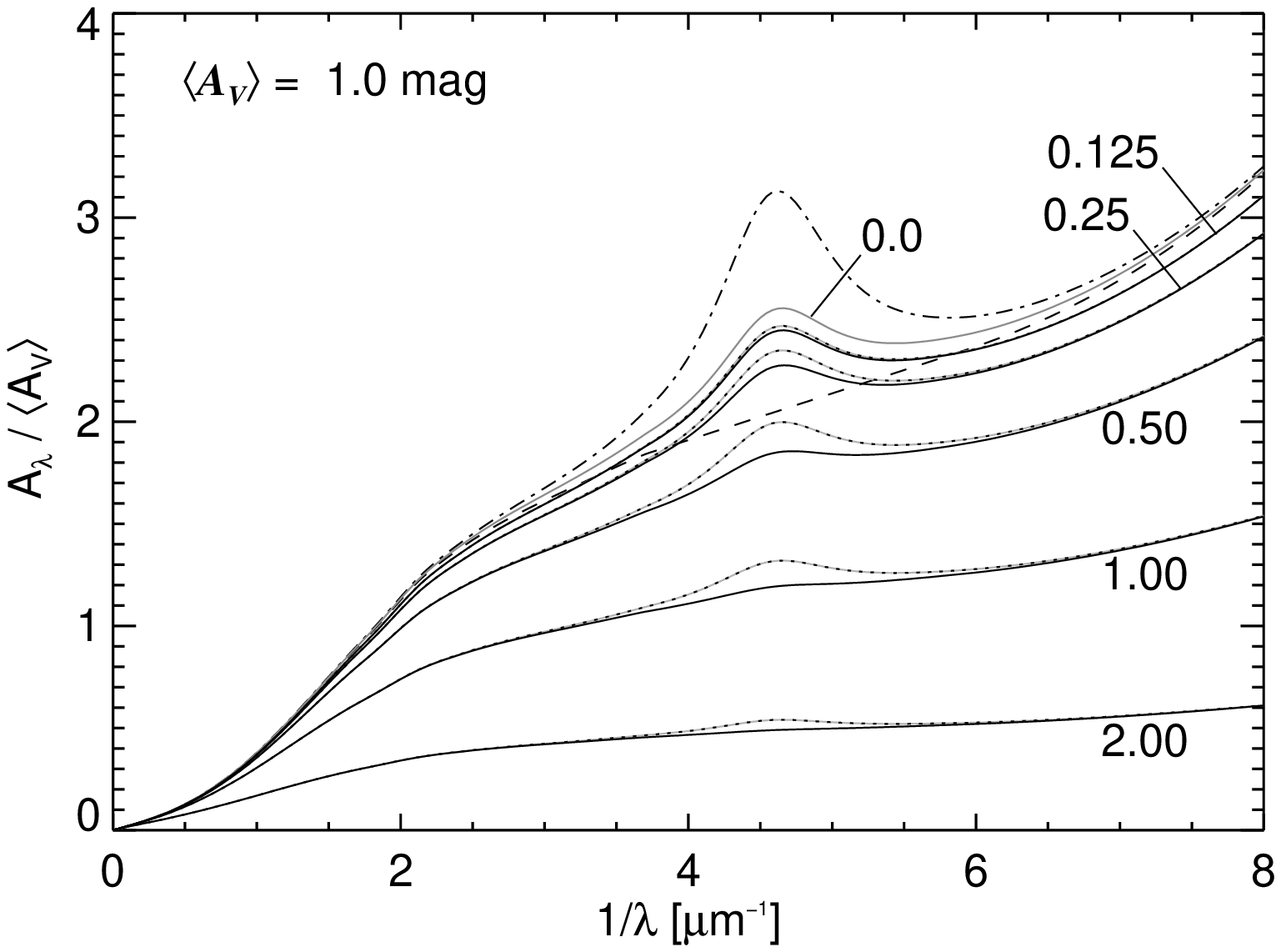}
	\hfill
	\includegraphics[width=0.49\hsize]{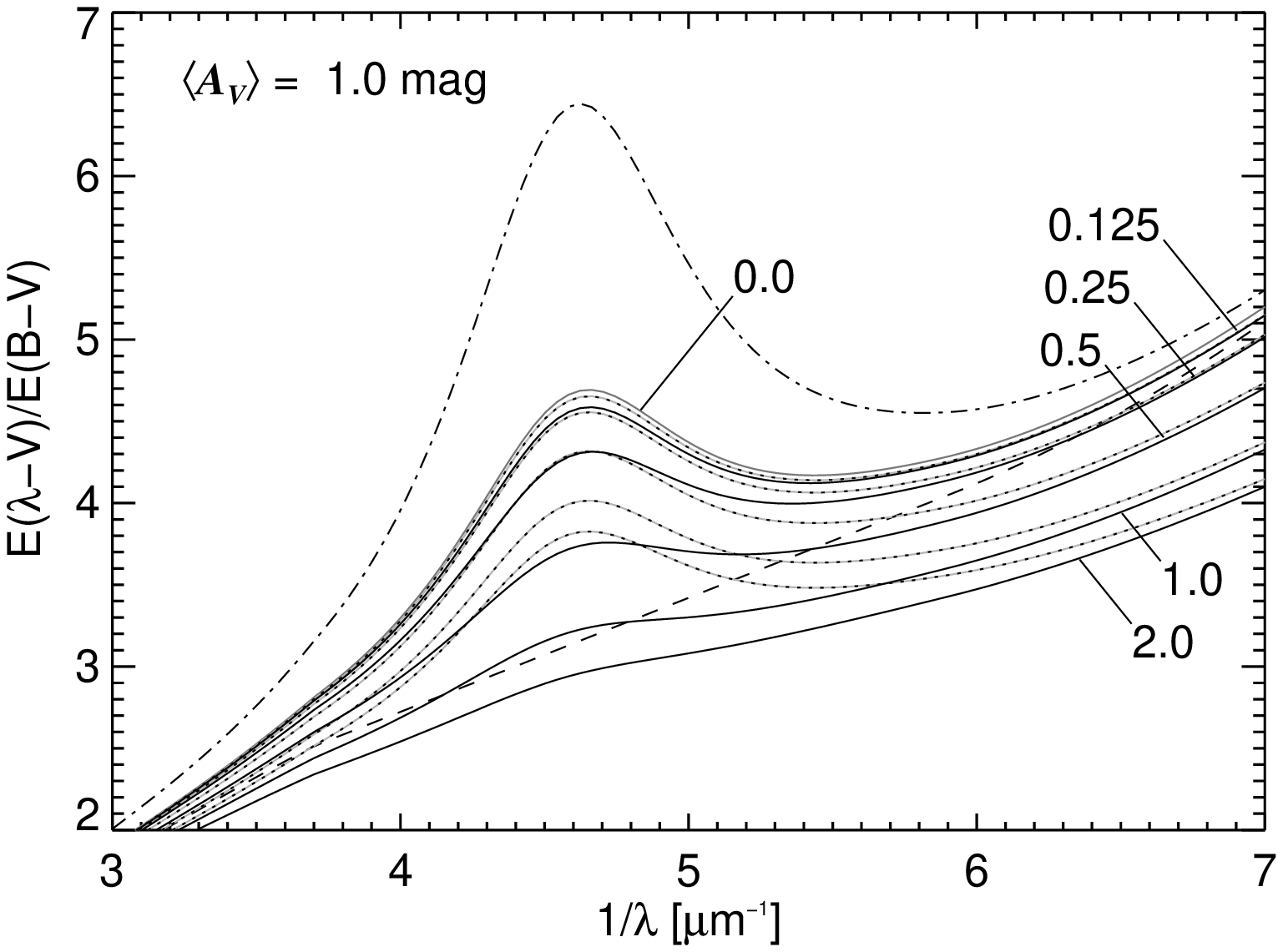}
	\includegraphics[width=0.49\hsize]{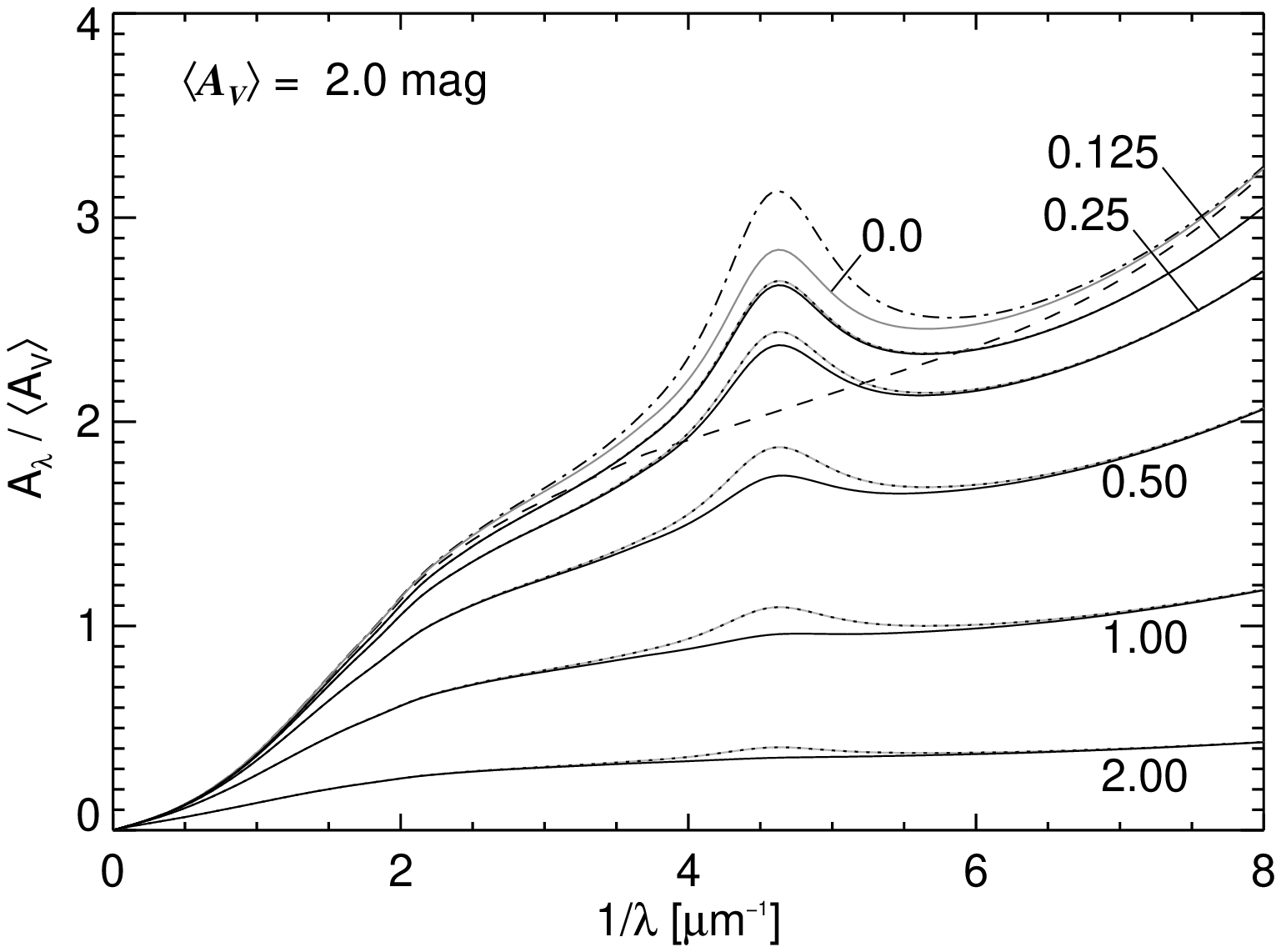}
	\hfill
	\includegraphics[width=0.49\hsize]{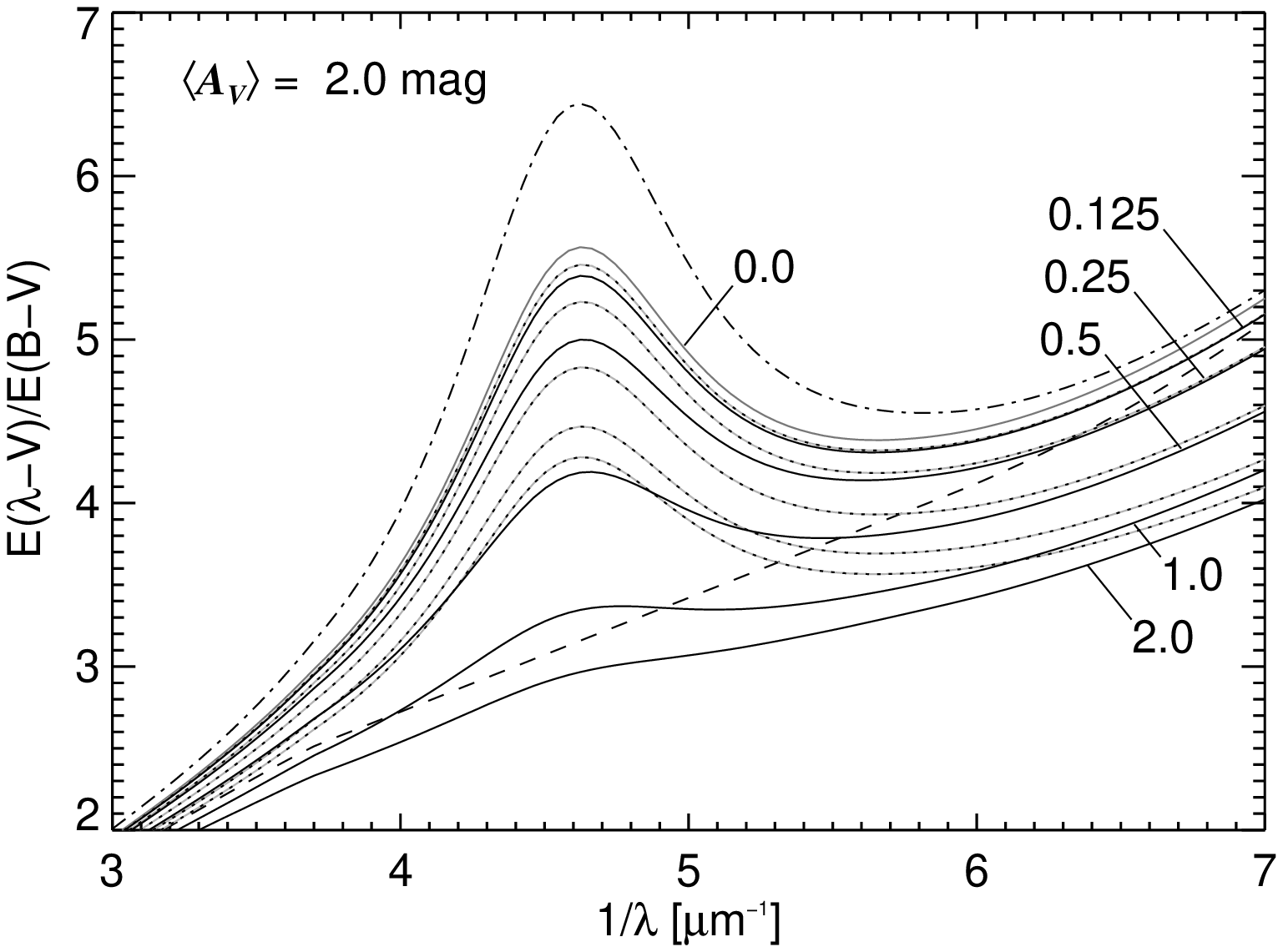}
	\caption{\label{fig_attcurves}
	Effective Extinction curves (left hand figure) and reddening curves around the extinction 
	bump (right hand figure) for a turbulent distant screen. 
	The intrinsic extinction curves with and without extinction bump are shown as dashed-dotted 
	and as dashed curve, respectively. 
	The effective extinction and reddening curves without and with change
	of the dust properties as function of optical depth are shown either as dotted curve or solid curves. 
	Added is also the mean extinction curve for the limit of a non turbulent screen (grey curve).
	It is assumed that
	the extinction bump at $2175~\AA{}$ is absent below a column density $N_{\rm H}=10^{21}~{\rm cm^{-2}}$
	which corresponds, assuming $R_V=3.1$ \citep{Fitzpatrick1999} and a dust-to-gas-ratio of 
	$N_{\rm H}/E(B-V)=5.8\times 10^{21}~{\rm cm^{-2}}$ \citep{Bohlin1978}, to $A_V\approx 0.53~{\rm mag}$.
	The curves are labelled with the corresponding standard deviation
	$\sigma_{\ln \xi}$ of the log-normal density distribution of the normalized column density 
	$\xi=N_{\rm H}/\left<N_{\rm H}\right>$.
	}
\end{figure*}

\subsection{The attenuation curve}

To derive attenuation curves we assume for the critical column density below which the carriers of the extinction
bump are destroyed $[N_{\rm H}]_{\rm crit}=10^{21}~{\rm cm^{-2}}$.
For the dusty screen we assumed several values of the standard deviation $\sigma_{\ln \xi}$ of the log-normal
distribution of the column density spanning a range from a smooth ($\sigma_{\ln \xi}=0.125$) to a highly non homogeneous
medium ($\sigma_{\ln\xi}=2$). To show the effect on the extinction value we considered two mean values $\left<A_V\right>=1~{\rm mag}$ and $\left<A_V\right>=2~{\rm mag}$. 

The derived curves are shown in Fig.~\ref{fig_attcurves}. The important parameters characterizing the curves 
are summarized in Tab.~\ref{table1} which are the effective absolute-to-relative extinction $R_V^A = A_V^{\rm eff}/E(B-V)^{\rm eff}$,
the extinction in V-band $A_V/\left<A_V\right>$, and the peak strength $c_3/\gamma^2$ which is analyzed more 
quantitatively in Sect.~\ref{sect_peakstrength}. In addition the table lists
the density contrast for certain assumptions for the power $n$ and the thickness of the screen relative to the maximum cloud size.
The density contrast for $\sigma_{\ln\xi}=2$ is already more than two times higher than is implied by CO measurements. 
In this regard the physical
conservative regime is limited to $\sigma_{\ln\xi}<2$. In systems with higher star formation rates like in star burst galaxies 
which lead to stronger turbulence the distributions are possibly wider.

As visualized in Fig.~\ref{fig_attcurves}, the distant screen becomes more transparent
for wider distributions of the column densities. This effect is stronger in more optically thick media.
In the optical this produces a flatter effective extinction curve which leads to a larger absolute to relative extinction 
$R_V^A$. As shown in Tab.~\ref{table1} for the considered parameters a broader distribution of the column density
produces a flatter extinction curve. However, we note that for extremely broad distributions this behavior is not
valid for opaque screens ($\left<A_V\right>>1~{\rm mag}$) as is shown in the next section.
As we have shown in our paper~I  that the turbulent distant screen model suggests
for the Calzetti extinction curve  an $R_V^A$-value larger
than 4. Our best solution provides $R_V^A\sim 4.75$. This implies $\sigma_{\ln\xi}>1$ for both extinction values.

To emphasize the effect of the carrier destruction of the 2175~\AA{} bump we considered also 
a turbulent medium where the carrier abundance is naturally lower. In case of a smooth
medium the two curves become identical. As can be seen in Fig.~\ref{fig_attcurves}, in case of a naturally lower carrier abundance
the peak strength only mildly decreases for broader distributions which keeps to be prominent feature in the effective extinction
curve. In contrast, if the abundance changes according to the column density because of destruction
the peak weakens strongly for wider column density distributions. For example,
for $\sigma_{\ln\xi}>1$ the peak strength is less than $20\%$ of the intrinsic value (Tab.~\ref{table1}). 
 
%
%
%
%

\begin{table}[htbp]
\caption{\label{table1}Parameters of the attenuation curve}
	\begin{tabular}{lc|ccc|ccc}
		& & \multicolumn{3}{c|}{$\left<A_V\right>=1~{\rm mag}$} & \multicolumn{3}{c}{$\left<A_V\right>=2~{\rm mag}$}\\
		$\sigma_{\ln\xi}$ & $\sigma_{\rho/\left<\rho\right>}$\tablefootmark{a} & $R_V^A$ & $\frac{A_V}{\left<A_V\right>}$ 
		& $c_3/\gamma^2$ & $R_V^A$ & $\frac{A_V}{\left<A_V\right>}$ & $c_3/\gamma^2$ \\
		\hline
		$0.125$	& 0.28  & 3.13 & 0.993 & 1.44 & 3.15 & 0.986 & 2.27\\
		$0.250$ & 0.57 & 3.21 & 0.972 & 1.21& 3.31 & 0.947 & 1.94 \\
		$0.500$ & 1.19 & 3.51 & 0.896 & 0.74 & 3.79 & 0.825 & 1.26 \\
		$1.000$ & 2.93 & 4.29 & 0.682 & 0.37 & 4.79 & 0.565 & 0.54 \\
		$2.000$ & 16.4 & 5.54 & 0.320 & 0.19 & 6.11 & 0.239 & 0.23 \\
	\end{tabular}	
	\tablefoottext{a}{Density contrast based on Eq.~\ref{eq_sigma} assuming $n=-10/3$ 
	and $\Delta/L_{\rm max}=1$.}
\end{table}


\subsection{Peak strength}

\label{sect_peakstrength}

\begin{figure*}[htbp]
	\includegraphics[width=0.49\hsize]{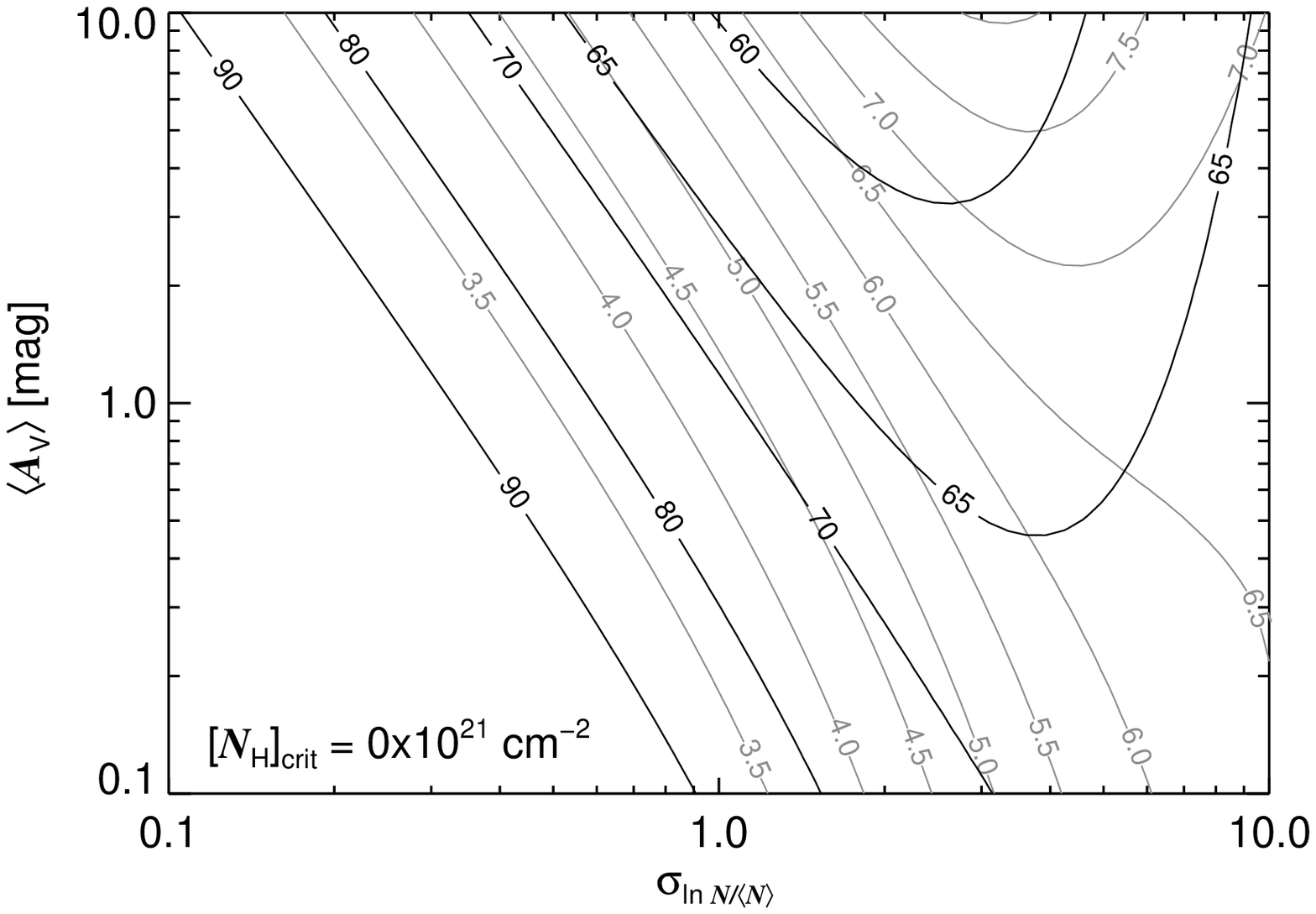}
	\hfill
	\includegraphics[width=0.49\hsize]{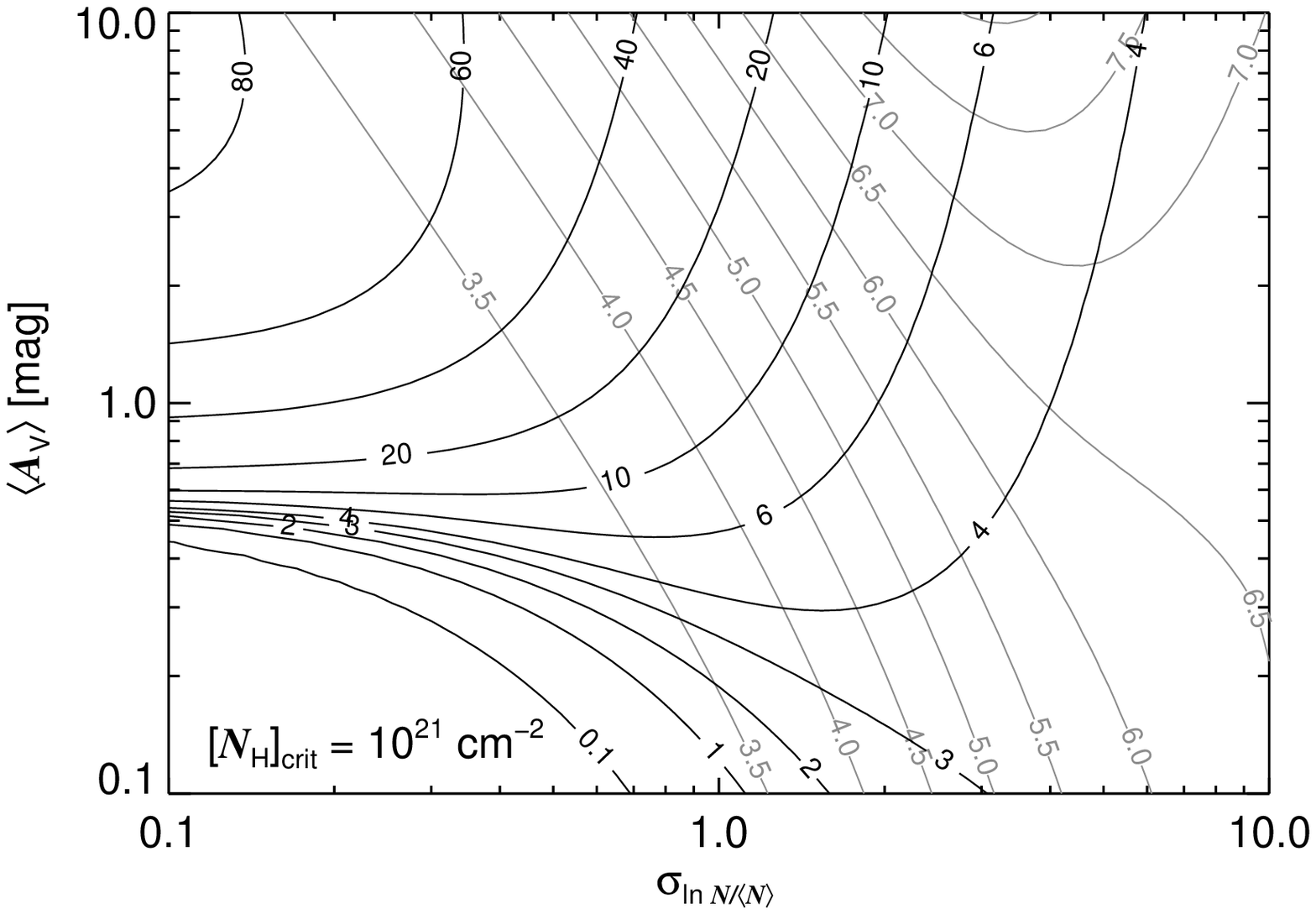}
	\caption{\label{fig_peakstrengtharr}
	Peak strength $\Delta A_{0.22}^{\rm eff}/E(B-V)^{\rm eff}$ of the effective extinction curve 
	without (left) and with additional destruction (right) of the carriers
	as function of the two parameters $\left<A_V\right>$ and $\sigma_{\ln N/\left<N\right>}$ of the turbulent
	screen. The black curves are lines of constant peak strengths. They are labelled with the percentage of the peak
	strength relative to the intrinsic value. 
	In the right figure the carriers are assumed to be destroyed below a critical column density of
	$[N_{\rm H}]_{\rm crit}=10^{21}~{\rm cm^{-2}}$. The corresponding $R_V$-values  
	are shown as grey solid lines.
	}
\end{figure*}

We have analyzed the effect of the additional destruction of a turbulent distant screen on the absorption
feature by considering the effective peak strength $\Delta A_{\rm 0.22}^{\rm eff}/E(B-V)^{\rm eff}$
where $\Delta A_{0.22}^{\rm eff}$ is the difference of the effective extinction
at peak frequency of the complete model and a model where the bump
has been removed from the intrinsic extinction curve.
To understand the importance of additional destruction we also considered
a screen where no further destruction has occurred.

\subsubsection{No additional destruction}

As found in paper~III the effective extinction curves of the turbulent screen are well determined 
by the effective absolute-to-relative extinction $R_V$. For given $R_V$-value we therefore expect
a certain strength of the peak.  As Fig.~\ref{fig_peakstrengtharr} shows, in case of probability distribution functions (PDFs)  
of the column densities with $\sigma_{\ln \xi}\ll 1$ this behavior is quite accurate. For wide
PDFs  and high extinction values the $R_V$-value only provides an approximation
of the correct attenuation curves. For example, if we consider a certain $R_V$-value
the peak strength increases for broader PDFs. But still, for 
$\sigma_{\ln \xi}<2$ the effect is in the order of only a few percent.

Turbulence not only flattens the extinction curve but also reduces the peak strength as we have seen 
in the former section. The effects are stronger 
in more optically thick media but, as Fig.~\ref{fig_peakstrengtharr} shows, do
not simply increase towards wider PDFs of the column density. This behavior is only
true for media which are optically thin and for optically thick media with not extremely wide PDFs. 
As shown in App.~\ref{app2} in the limit of infinitely broad 
PDFs the $R_V$-value and the peak strength of the effective extinction curve become independent
on the mean extinction $\left<A_V\right>$ and the standard deviation of the column density $\sigma_{\ln\xi}$.
The asymptotic absolute-to-relative extinction is given by $R_V^{\rm eff}=\sqrt{R_V}/(\sqrt{R_V+1}-\sqrt{R_V})$. 
For $R_V=3.1$
we have $R_V^{\rm eff}\approx 6.67$. Likewise, we have a limit of the peak strength given by 68\%. 
For media which are optically thick the asymptotic value does not provide the strongest effect on the
flatness and peaks strength. But still,
for the considered parameter range the peak strength is quite strong with $>50\%$. 
Turbulence alone is therefore not able to produce the low peak strength of the Calzetti curve.

\subsubsection{Additional destruction}

\begin{figure*}[htbp]
	\includegraphics[width=0.49\hsize]{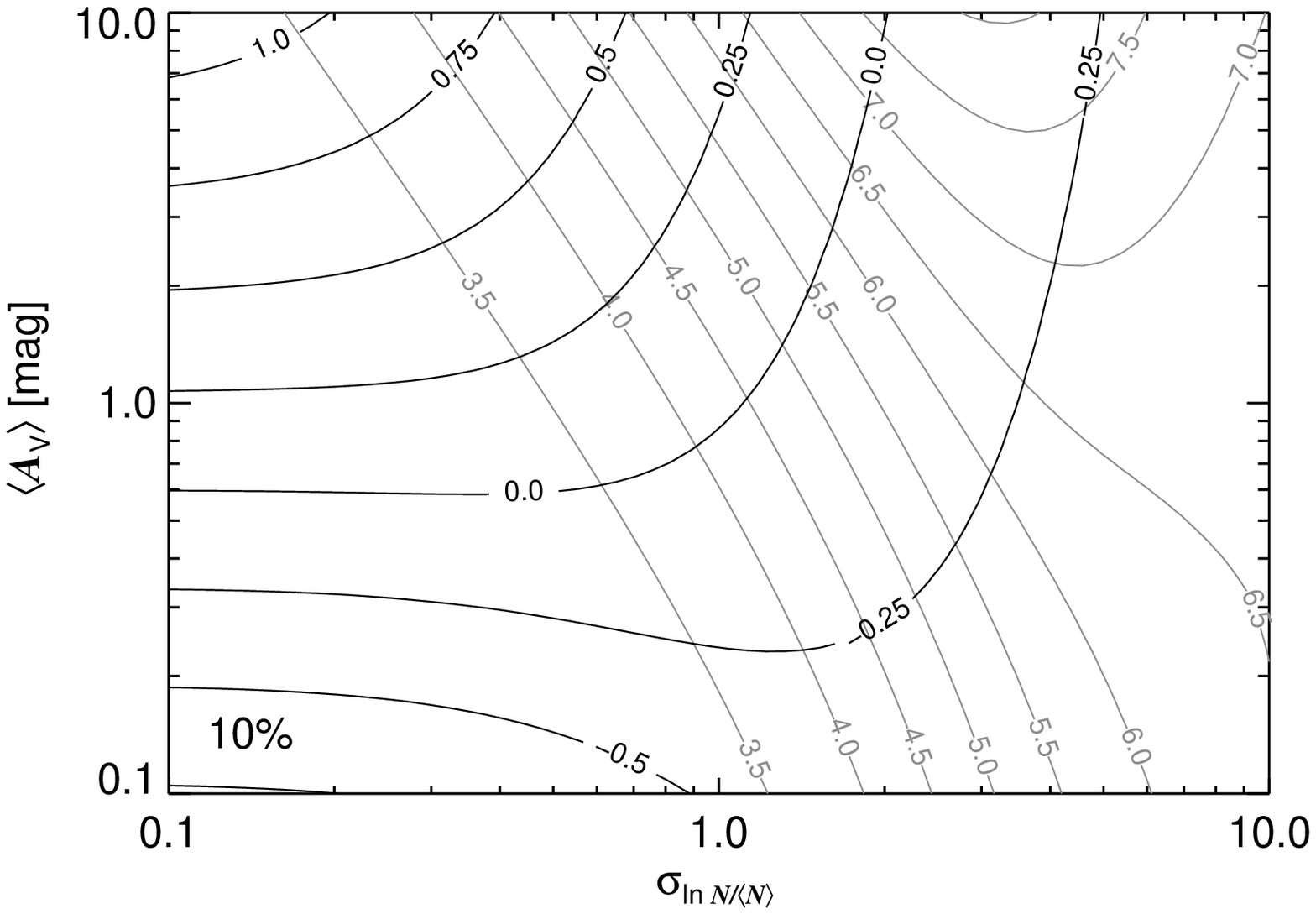}
	\hfill
	\includegraphics[width=0.49\hsize]{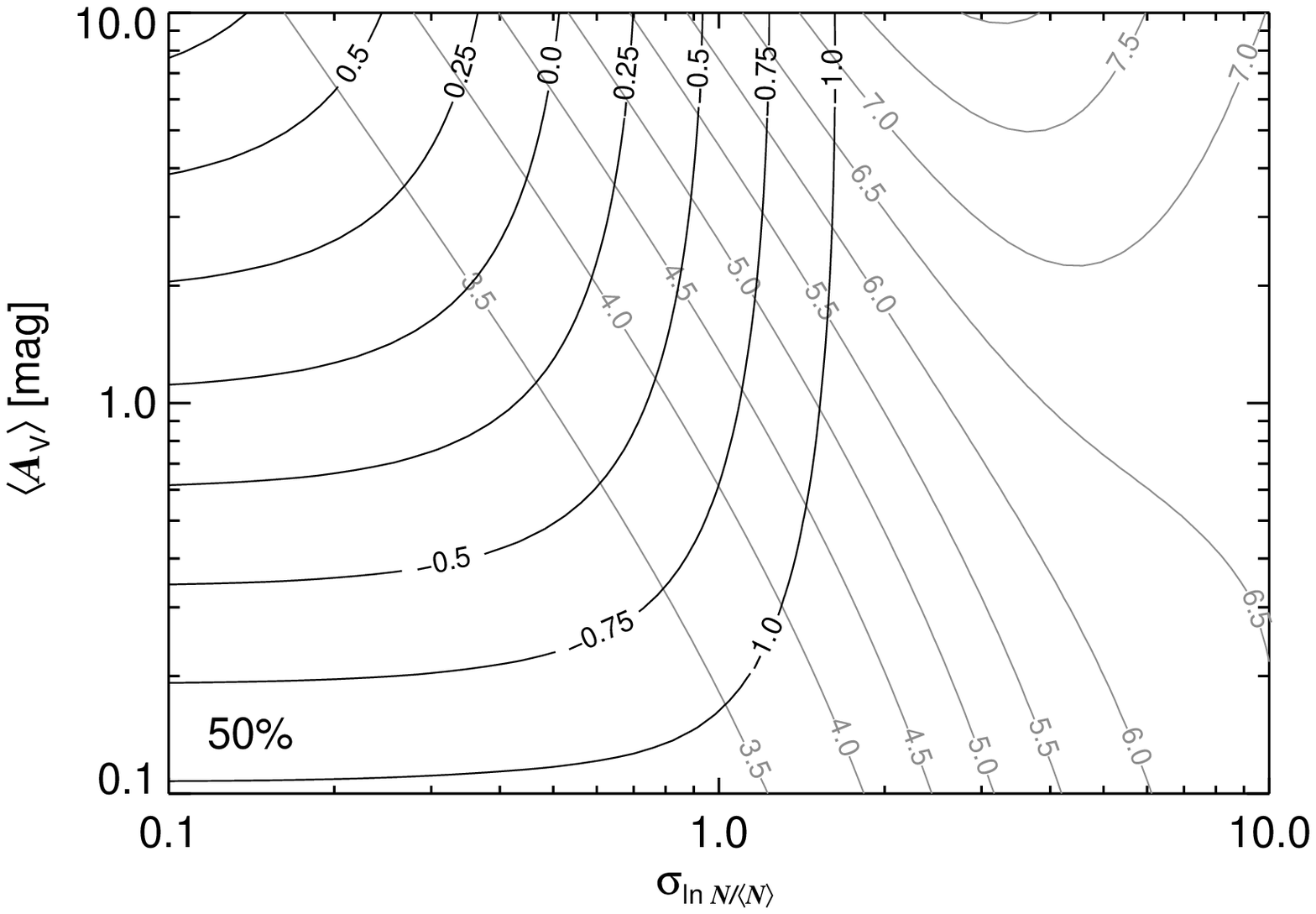}
	\caption{\label{fig_critcol}
	Critical column density $[N_{\rm H}]_{\rm crit}$ needed to reduce the peak strength 
	$\Delta A^{\rm eff}_{0.22}/E(B-V)^{\rm eff}$ of the effective attenuation curve
	of a turbulent screen to 10\% (left hand figure) or 50\% (right hand figure) of the intrinsic value. 
	The curves are labelled
	by $\log N_{\rm H}[10^{21}~{\rm cm^{-2}}]$. The grey curves give the effective $R_V$-value.
	}
\end{figure*}

As a special example to analyze the effect caused by the additional destruction of the carriers of the peak on
its strength we considered again a critical column density of $[N_{\rm H}]_{\rm crit}=10^{21}~{\rm cm^{-2}}$.
Fig.~\ref{fig_peakstrengtharr} shows a strong reduction of the peak strength even for less broad PDFs. For
mean column densities well above the critical column densities the peak strength weakens strongly in case of
more turbulent media. For $\left<A_V\right><10~{\rm mag}$ and $\sigma_{\ln \xi}>2$ the peak strength is 
lower than 10\% relative to the intrinsic value. The impact of the destruction on the carriers
weakens for higher extinction values $\left<A_V\right>$. 

In case of turbulent screens with mean column densities well below the critical column density turbulence 
produces regions of high column densities where the carriers of the peak can survive. As the mass is compressed to
more opaque clouds in higher turbulent media the peak strength increases with $\sigma_{\ln \xi}$.

For intermediate mean column densities turbulence leads to an increase at low $\sigma_{\ln \xi}$ but to a decrease
of the peak strength at high $\sigma_{\ln \xi}$. 

In the limit of broad PDFs of the column density the peak strength reaches asymptotically a value which is independent
on the main parameters of the screen, the mean extinction $\left<A_V\right>$ and standard deviation of the log-normal
function $\sigma_{\ln \xi}$. It solely depends on the critical column density $[N_{\rm H}]_{\rm crit}$ and allows therefore
a first estimate of the possible effect caused by the additional destruction on the peak strength. The asymptotic behavior
is analyzed in App.~\ref{app2}. For the critical column density assumed in Fig.~\ref{fig_peakstrengtharr} the asymptotic
value is 3.3\% of the intrinsic peak strength. As shown in App.~\ref{app2} for larger critical column densities 
the asymptotic value of the peak strength decreases strongly as 
$\tau_1^{-3/2}\,e^{-\tau_1}$ where $\tau_1=\kappa_{0.22}^{(1)}[N_{\rm H}]_{\rm crit}$.
For critical column densities well below $[N_{\rm H}]_{\rm crit}=10^{21}~{\rm cm^{-2}}$ the asymptotic peak strength reaches the
value of $68\%$ caused by the turbulence in the limit of broad PDFs as discussed above.

The effect of the critical column density on the peak strength is more accurately derived in Fig.~\ref{fig_critcol}.
The figure shows the critical column density 
needed to reduce the peak strength to 10\% and 50\% of
the intrinsic value. To produce a peak strength of turbulent screens with $\sigma_{\ln\xi}>1$ and $\left<A_V\right><10~{\rm mag}$ 
by more than $50\%$ the critical column density needs to be at least $\sim 3\times 10^{20}~{\rm cm^{-2}}$.
To decrease in the same parameter range of the screen the peak strength to 10\% of its intrinsic value 
the critical column density needs to be at least $\sim 2\times 10^{21}~{\rm cm^{-2}}$. As the figure shows 
the critical column density cannot be considerably smaller than $10^{21}~{\rm cm^{-2}}$ to produce peak strengths
as low as 10\%. For example, a critical column density of $6\times 10^{21}~{\rm cm^{-2}}$ would require in case
of screens ($\left<A_V\right>>1~{\rm mag}$) very broad PDFs with $\sigma_{\ln\xi}>4$ which would imply
Mach numbers $M>1500$ far above the ones measured in the ISM of our galaxy.

\section{Conclusion}

The analysis shows that a turbulent distant screen can naturally explain not only the flatter curvature
but also a weak absorption feature at 2175~\AA{} of the Calzetti-curve if within a certain column density
its carriers are destroyed by the strong UV-radiation in star-burst galaxies.
The feature is efficiently reduced by more than 80\% of its intrinsic value if the following circumstances are fulfilled for typical extinction values measured for star burst galaxies ($A_V\sim 1~{\rm mag}$):
\begin{itemize}
	\item The standard deviation of the log-normal distribution must be larger than $\sigma_{\ln\xi}=1$.
	\item The critical column density needs to be larger than $[N_{\rm H}]_{\rm crit} >10^{21}~{\rm cm^{-2}}$.
\end{itemize}
The condition for the standard deviation of the log-normal distribution of the column density is in agreement with the 
Mach number of the cold neutral medium (which implies $\sigma_{\rho/\left<\rho\right>}\sim 6$) if the thickness is not sufficiently larger than a few turbulent length scales.

A fractal density structure can also enhance the probability for the carriers' survival as
they become located in optical thick clouds and therefore save against further destruction by a strong UV-field.

\begin{acknowledgements}
	Dopita \& Fischera acknowledge financial support under Discovery project DP0984657.
\end{acknowledgements}

\bibliographystyle{aa} 
\bibliography{16644reference}

\appendix

\section{Approximation}
\label{app}
If we consider an optically thick medium then turbulence will lead to regions of low column 
density through which most of the light will be transmitted. The effective attenuation curve is
determined by those regions. If the medium becomes on the other hand highly turbulent the medium is compressed
to very small clouds. The only attenuation occurs in regions of high column density which determine
the attenuation curve. To differentiate the two different cases we can consider the column density 
$N_{\rm H} = \left<N_{\rm H}\right>e^y$ where $\tau = 1$
relative to the column density at maximum position of the log-normal distribution 
$N_{\rm H} = \left<N_{\rm H}\right>e^{-0.5\sigma^2_{\ln \xi}}$.
We distinguish the cases $-\ln \left<\tau\right> \ll -0.5\sigma^2_{\ln \xi}$ and
$-\ln \left<\tau\right> \gg -0.5\sigma^2_{\ln \xi}$

\begin{figure*}[htbp]
	\includegraphics[width=0.49\hsize]{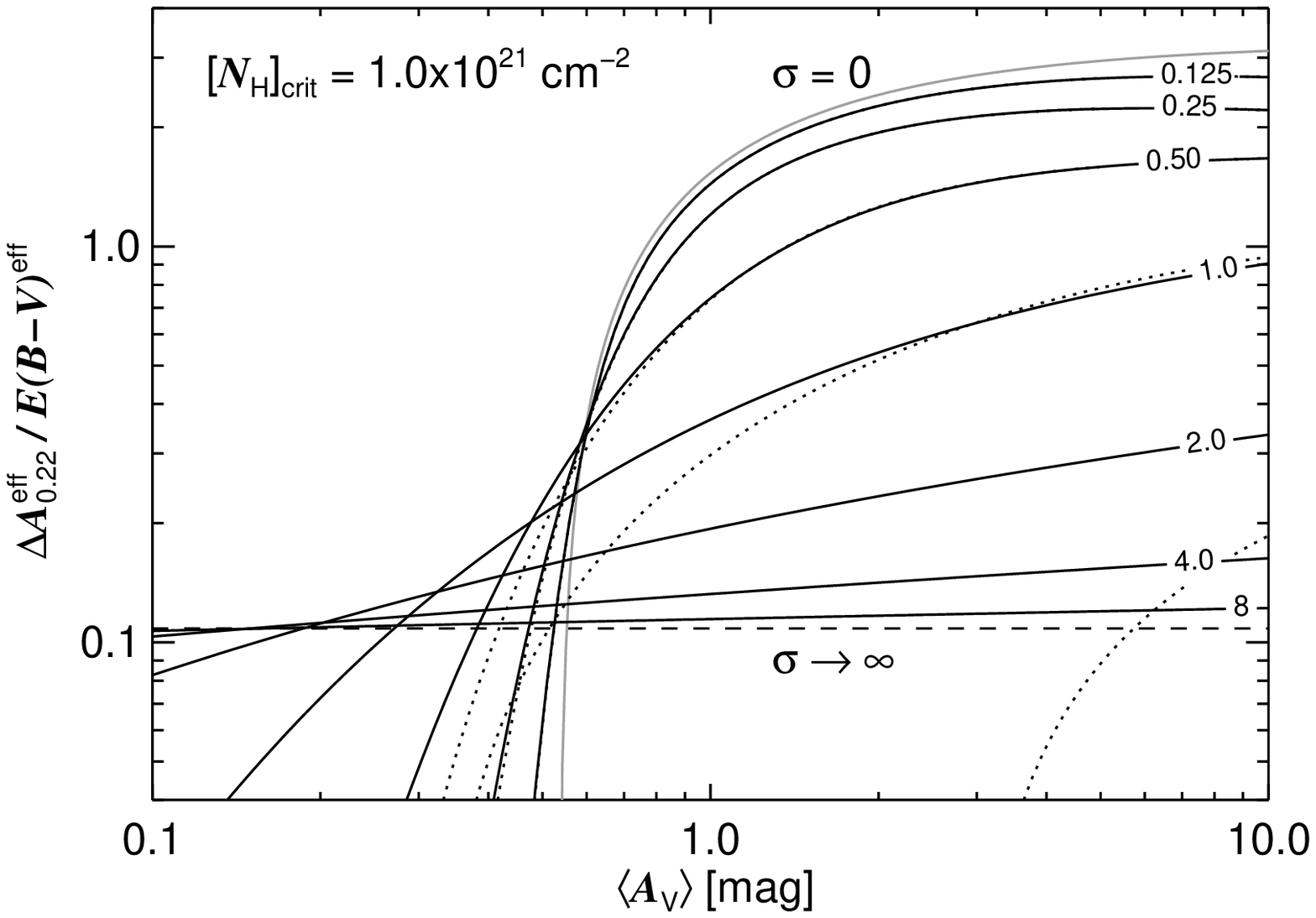}
	\hfill
	\includegraphics[width=0.49\hsize]{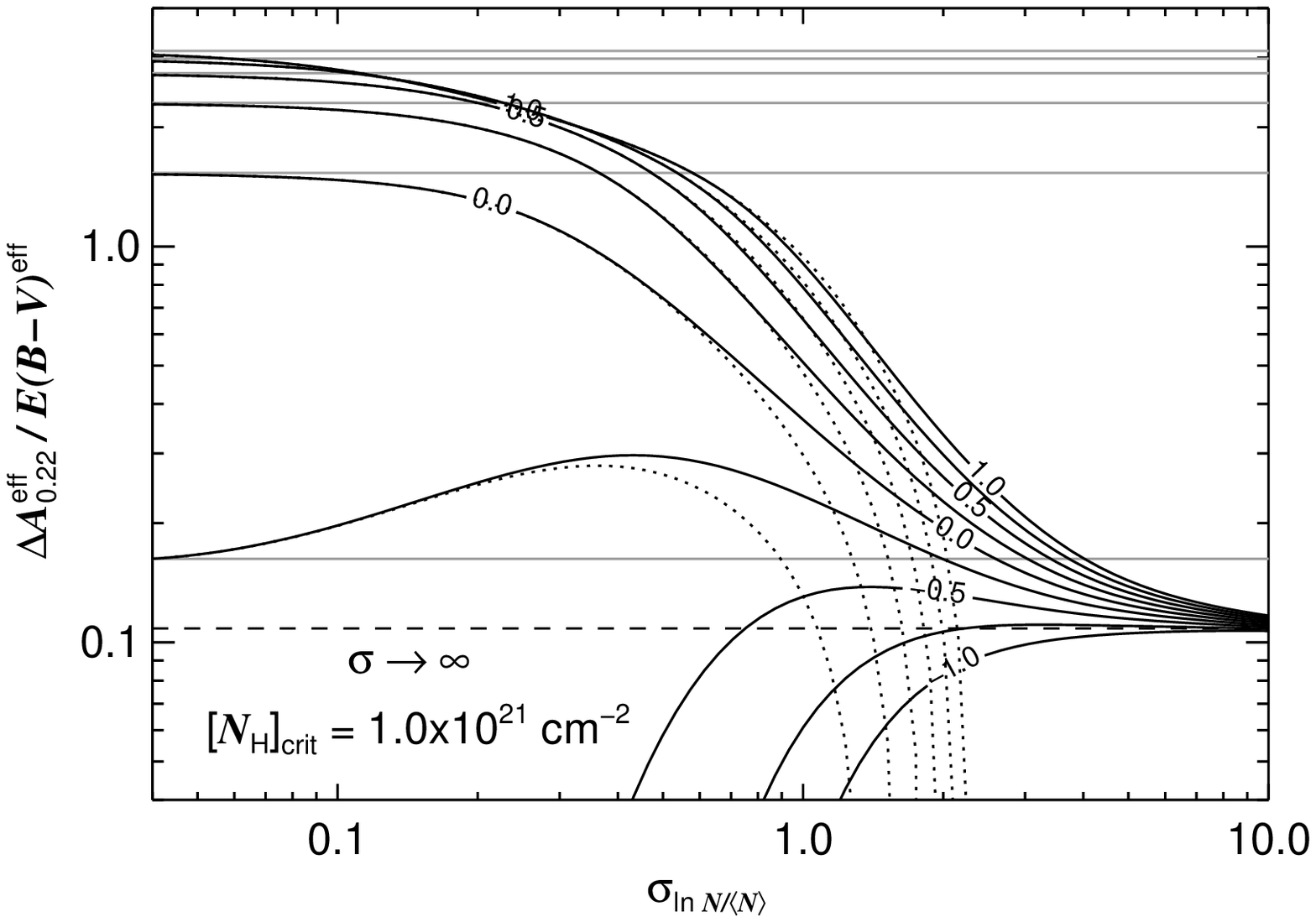}
	\caption{\label{fig_peakstrength_cuts}
	Peak strength of the effective extinction curve caused by a distant turbulent screen as function 
	of the mean extinction
	($\left<A_V\right>$) for given column density contrast ($\sigma_{ln N/\left<N\right>}$) (left hand figure)  and 
	as function of column density contrast for given mean extinction (right hand figure). The curves in the right
	hand figure are labelled with $\log A_V [{\rm mag}]$.
	The critical column density is assumed to be $[N_{\rm H}]_{\rm crit}=10^{21}~{\rm cm^{-2}}$. 
	The values in the limit of a non turbulent medium ($\sigma=0$) are shown as grey lines, the value
	in the the theoretical limit of infinite column density contrast ($\sigma\rightarrow \infty$) as dashed line.
	Also shown as dotted lines are peak strengths derived using the approximation Eq.~\ref{eq_efftaumodel}.}
\end{figure*}

\subsection{Approximation for $\ln\left<\tau\right>\gg 0.5\sigma^2_{\ln\xi}$}
\label{app1}

The deviation of the approximation for $\ln\left<\tau\right>\gg 0.5\sigma^2_{\ln\xi}$ follows the procedure presented in paper~III. For sake of simplicity we ignore the additional dependence on wavelength. In this limit  the integrands of 
Eq.~\ref{eq_taueff} become narrow functions around the maxima at $\tilde y_i$ determined by
\begin{equation}
	f'(\tilde y_i) = -\left<\tau_i\right>e^{\tilde y_i} - \frac{1}{\sigma^2}(\tilde y_i +\frac{1}{2}\sigma_{\ln\xi}^2) = 0
\end{equation}
where $\tilde y_1$ is the location of the maximum of the first and $\tilde y_2$ the location of the maximum of the second integrand.
Developing the exponential function in the exponent around these maxima to a secondary order polynomial
funtion
\begin{equation}
	e^{y} \approx e^{\tilde y_i}(1+(y-\tilde y_i)+\frac{1}{2}(y-\tilde y_i)^2)
\end{equation}
for $i=1,2$ leads to the approximate expression of the effective extinction
\begin{eqnarray}
	\label{eq_efftaumodel}
	e^{-\tau^{\rm eff}} &=& e^{-\tau_1^{\rm eff}}\frac{1+erf(t_1)}{2}
	+e^{-\tau_2^{\rm eff}}\frac{1-erf(t_2)}{2}.
\end{eqnarray}
where
\begin{equation}
	erf(t_i) = \frac{2}{\sqrt{\pi}}\int_0^{t_i}{\rm d}y\,e^{-y^2}
\end{equation}
is the error function and where
\begin{equation}
	t_i = \sqrt{\frac{\gamma_i}{2\sigma_{\ln\xi}^2}}(y_{\rm crit}-\tilde y_i), \quad \gamma_i = 1+\left<\tau_i\right>\sigma_{\ln\xi}^2 e^{\tilde y_i}.
\end{equation}
The effective optical depths are given by:
\begin{eqnarray}
	\tau_1^{\rm eff} & = & \frac{1}{2}\ln\gamma_1+\frac{1}{2}\left<\tau_1\right>e^{\tilde y_1}(1+\gamma_1),\\
	\tau_2^{\rm eff} & = & \frac{1}{2}\ln\gamma_2+\Delta\left<\tau\right>e^{y_{\rm crit}}+\frac{1}{2}\left<\tau_2\right>e^{\tilde y_2}(1+\gamma_2).
\end{eqnarray}
In paper~III we have shown that the approximation provides the correct results also for small fluctuations $\sigma_{\ln \xi}\ll 1$ for all optical depths. In the limit of small fluctuations we obtain the natural optical depths of a 
homogeneous screen as $\gamma_i\rightarrow 1$ and $\tilde y_i\rightarrow 0$.

Fig~\ref{fig_peakstrength_cuts} shows peak strengths derived using the approximation. They are compared with accurate 
calculations. The approximation becomes more accurate for
smaller standard deviations and larger extinction values. 
For optical thick media ($\left<A_V\right>>1~{\rm mag}$) the 
approximation is accurate for $\sigma_{\ln\xi}\le 1$.

The approximation can be used to estimate if the effective extinction curve is either a flat curve, so that
$\tau^{\rm eff}_\lambda\approx \tau^{\rm eff,1}_\lambda$, or `peak dominated', so that $\tau_\lambda^{\rm eff}\approx\tau_\lambda^{\rm eff,2}$. The effective extinction is 'peak dominated' if the critical value lies well below the typical 
column densities contributing to the effective extinction $e^{-\tau^{\rm eff,2}}$. As a simple criterium we can consider
$t_i=0$ for $i=1,2$ which provides for given mean and critical column density a critical standard deviation
\begin{equation}
	\sigma_{\ln \xi}^{\rm crit,i} = \sqrt{\frac{-2 y_{\rm crit}}{2\kappa_{0.22~\mu{\rm m}}^{(i)}[N_{\rm H}]_{\rm crit}+1}},\quad y_{\rm crit}\le 0.
\end{equation}
where $\kappa_{0.22\,\mu{\rm m}}^{(i)}$ are the extinction coefficients given in Eq.~\ref{eq_extcoeffpeak}.

The extinction curve is `peak dominated' for $t_2\ll 0$ ($\sigma_{\ln\xi}\ll \sigma_{\ln\xi}^{\rm crit,2}$) as the error functions become $erf(t_1)=erf(t_2)=-1$.
In the additional limit $[N_{\rm H}]_{\rm crit}/\left<N_{\rm H}\right>\ll 1$ we obtain the result of paper~III
\begin{equation}
	\tau_\lambda^{\rm eff,2}  =  \frac{1}{2}\ln\gamma_2+\frac{1}{2}\left<\tau^2\right>e^{\tilde y_2}(1+\gamma_2).
\end{equation}
This solution is trivial for $\kappa_\lambda^{(1)}=\kappa_\lambda^{(2)}$ as $t_1=t_2$. The effective extinction 
becomes essentially flat for $t_1\gg 0$ ($\sigma_{\ln\xi}\gg\sigma_{\ln\xi}^{\rm crit,1}$). A number of critical
standard deviations are listed for given mean extinction and critical column density in Tab.~\ref{table_critsigma}.
The values point to a critical column density $[N_{\rm H}]_{\rm crit}\ge 10^{21}~{\rm cm^{-2}}$ to obtain
a flat extinction curve as lower values would imply wider PDFs which become unlikely considering the
density contrast in the ISM of our Galaxy.

\begin{table}[htbp]
	\caption{\label{table_critsigma}Critical standard deviations}
	\begin{tabular}{lc|cccc}
	$\left<A_V\right>$ & [mag] & 1.00 & 2.00 & 1.00  & 2.00 \\
	$[N_{\rm H}]_{\rm crit}$ & $[10^{21}~{\rm cm^{2}}]$ &  0.50 & 0.50 &  1.00 & 1.00 \\
	\hline
	$\sigma_{\ln \xi}^{\rm crit,1}$ 	&  & 1.15 & 1.42 &  0.65 & 0.94 \\
	$\sigma_{\ln \xi}^{\rm crit,2}$ 	& & 1.02 & 1.27 &  0.56 & 0.81 \\
 	\end{tabular}	
\end{table}

\subsection{Approximation for $\ln\left<\tau\right>\ll 0.5\sigma^2_{\ln \xi}$}
\label{app2}

In the limit $-\ln \left<\tau\right> \gg -0.5\sigma^2_{\ln \xi}$ most sight lines through the turbulent
medium are optically thin. It is therefore convenient to rewrite the equation for the effective optical depth
so that
\begin{equation}
	\label{eq_taueffmod}
	\tau_\lambda^{\rm eff} = -\ln\left\{1-\int{\rm d}y\,p(y)\,\left(1-e^{-\left<\tau\right>e^{y}}\right)\right\}.
\end{equation}
The integral is a small number so that
\begin{equation}
	\label{eq_taueffmod2}
	\tau_\lambda^{\rm eff} \approx \int{\rm d}y\,p(y)\,\left(1-e^{-\left<\tau\right>e^{y}}\right).
\end{equation}

In case that $y\ll 0.5\sigma^2$ we can replace the PDF of the column density through a power law distribution:
\begin{equation}
	\label{eq_powerlawdistr}
	{\rm d}y\,p(y) \approx d\tau\frac{ e^{-\sigma^2/8} }{\sqrt{2\pi}\sigma}\frac{\sqrt{\left<\tau\right>}}{\sqrt[3]{\tau}}
\end{equation}
where $\tau= \left<\tau\right>\ln y =\kappa[N_{\rm H}]_{\rm crit}$.
The equation for the effective optical depth becomes after an additional partial integration:
\begin{equation}
	\tau^{\rm eff} \approx \sqrt{\frac{2}{\pi}}\frac{e^{-\sigma^2/8}}{\sigma}\sqrt{\left<\tau\right>} \int_0^{\infty}{\rm d}t\,t^{-1/2}\,e^{-t}
\end{equation}
where the integral can be identified as the $\Gamma$-function with $\Gamma(\frac{1}{2})=\sqrt{\pi}$.

In the limit of a broad PDF the effective optical depth decreases strongly with $\sigma_{\ln \xi}$ and is proportional to
the square root of the mean optical depth. The ratio of two effective optical depths becomes not only independent on the 
width of the PDF but also independent on the dust content of the screen.
As a special case we can consider the absolute to relative extinction value $R_V=A_V/(A_B-A_V)$. The corresponding
value of a turbulent screen with an infinite broad PDF has then a value of $1/R_V^{\rm eff}=\sqrt{1+R_V^{-1}}-1$. For
$R_V=3.1$ we obtain therefore an asymptote $R_V^{\rm eff}\approx6.67$. 
Another example is the peak strength. If we ignore additional destruction the peak strength in case of broad PDFs
of the column density approaches asymptotically
\begin{equation}
	\frac{\Delta A^{\rm eff}}{E(B-V)^{\rm eff}} = \frac{\sqrt{\kappa_{0.22}^{(2)}}-\sqrt{\kappa_{0.22}^{(1)}}}{\sqrt{\kappa_B}-\sqrt{\kappa_V}}=2.23.
\end{equation}
Compared to the intrinsic value of $3.30$ this means a reduction to 68\% in the limit of infinite broad PDFs.

%

A similar approach leads to the asymptotic behavior of the peak strength in case of additional destruction. The equation
for the effective optical depth at peak frequency can be rewritten as:
\begin{eqnarray}
	\tau^{\rm eff} =-\ln&\Big\{&1-\int_{y_{\rm crit}}^{\infty}{\rm d}y\,p(y)\,\left(1-e^{-\Delta \left<\tau\right>_{\rm crit}}\right)
			\nonumber\\
		&- &  \int_{-\infty}^{\infty}{\rm d}y\,p(y)\, \left(1-e^{-\left<\tau_1\right>e^{y}}\right)\nonumber\\
		&+& \int_{y_{\rm crit}}^{\infty}{\rm d}y\,p(y)\,\left(1-e^{-\left<\tau_1\right>e^{y}}\right)\nonumber\\
		&-& e^{-\Delta\left<\tau\right>_{\rm crit}}\int_{y_{\rm crit}}^{\infty}{\rm d}\,p(y)\,\left(1-e^{-\left<\tau_2\right>e^{y}}\right)\Big\}.
\end{eqnarray}
The sum over the four integrals is a small number so that we can make the same simplification as in Eq.~\ref{eq_taueffmod2}.
Replacing the distribution of the column density by the power law distribution (Eq.~\ref{eq_powerlawdistr})
provides:
\begin{eqnarray}
	\tau^{\rm eff} \approx \sqrt{\frac{2}{\pi}}\frac{e^{-\sigma^2/8}}{\sigma}&\Big\{&
	\sqrt{\left<N_{\rm H}\right>/[N_{\rm H}]_{\rm crit}}\left(1-e^{-\Delta\left<\tau\right>_{\rm crit}}\right)\nonumber\\
	&&+\sqrt{\left<\tau_1\right>}\,(f(0)-f(\tau_1))\nonumber\\
	&&+\sqrt{\left<\tau_2\right>}\,f(\tau_2)\,e^{-\Delta\left<\tau\right>_{\rm crit}}\Big\},
\end{eqnarray}
where 
\begin{eqnarray}
	f(\tau) &=& \frac{1}{2}\int_\tau^{\infty}{\rm d}t\,t^{-3/2}\left(1-e^{-t}\right) \nonumber\\
		&=& \frac{1-e^{-\tau}}{\sqrt{\tau}}+\Gamma\left(\frac{1}{2},\tau\right).
\end{eqnarray}
$\Gamma(a,x)$ is the incomplete $\Gamma$-function
\begin{equation}
	\Gamma(a,x) = \int_x^\infty{\rm d}t\,t^{a-1}e^{-t}.
\end{equation}
For the special case $\tau=0$ we have $f(0)=\sqrt{\pi}$.

\begin{figure}[htbp]
	\includegraphics[width=\hsize]{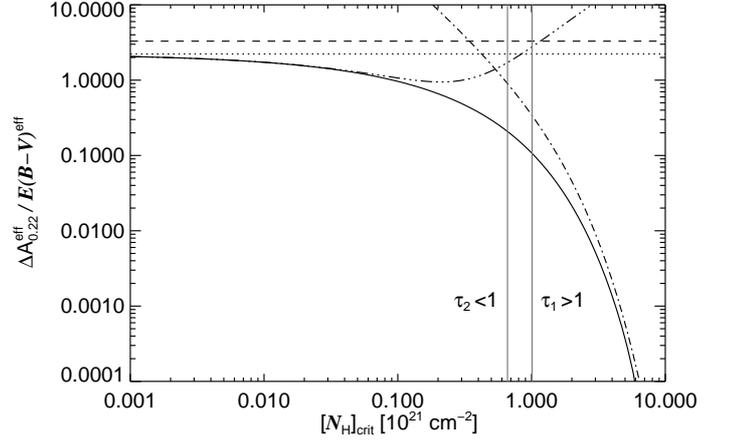}
	\caption{\label{fig_peakstrengthlimit}
	Peak strength in the limit of a broad PDF with $\ln\left<\tau\right>\ll \frac{1}{2}\sigma^2_{\ln\xi}$ as function
	of the critical column density below which the carriers of the peak are assumed to be destroyed.
	The intrinsic peak strength as given by \cite{Fitzpatrick1999} is shown as dashed line. The
	corresponding peak strength of a turbulent screen model with no additional destruction is shown as dotted 
	line. The dashed-dotted and the three dashed-dotted curves are the approximations in the regime $\tau_1=\kappa_{0.22}^{(1)}[N_{\rm H}]_{\rm crit}\gg 1$ and $\tau_2 = \kappa_{0.22}^{(2)}[N_{\rm H}]_{\rm crit}\ll 1$ given in Eq.~\ref{eq_approxhigh} and \ref{eq_approxlow}.}
\end{figure}

For the peak strength in the limit of an infinite broad PDF we obtain:
\begin{eqnarray}
	\frac{\Delta A^{\rm eff}_{0.22}}{E(B-V)^{\rm eff}} = && \frac{1}{\sqrt{\pi}\sqrt{\kappa_B}-\sqrt{\pi}\sqrt{\kappa_V}}\nonumber\\
	&&\times \Big\{\frac{1}{\sqrt{[N_{\rm H}]_{\rm crit}}}(1-e^{-\Delta\tau_{\rm crit}})\nonumber\\
	&&\quad+\sqrt{\kappa_2}\,f(\tau_2)\,e^{-\Delta\tau_{\rm crit}}-\sqrt{\kappa_1}\,f(\tau_1)\Big\}
\end{eqnarray}
The curve is shown in Fig.~\ref{fig_peakstrengthlimit}.
For $[N_{\rm H}]_{\rm crit}\ll 10^{21}~{\rm cm^{-2}}$ 
the asymptotic value becomes equal to the peak strength with no further destruction. 
The asymptotic value for $[N_{\rm H}]_{\rm crit}=10^{21}~{\rm cm^{-2}}$ is  added in Fig.~\ref{fig_peakstrength_cuts}. 
In the limit $y_{\rm crit}\gg -\ln \left<\tau_1\right>$ we can replace the incomplete Gamma function by
\begin{equation}
	\int_x^\infty t^{a-1} \,e^{-t} \approx x^{s-1}e^{-x}\left(1+(s-1)x^{-1}\right)
\end{equation}
which leads to 
\begin{eqnarray}
	\label{eq_approxhigh}
	\Delta\tau^{\rm eff}  &\approx &\sqrt{\frac{2}{\pi}}\frac{e^{-\sigma^2/8}}{\sigma}\nonumber\\
		&&\times \sqrt{\frac{\left<N_{\rm H}\right>}{[N_{\rm H}]_{\rm crit}}}\frac{1}{2}\left(
			\frac{1}{\tau_1}-\frac{1}{\tau_2}\right)e^{-\tau_1}
\end{eqnarray}
At $[N_{\rm H}]_{\rm crit}\gg 10^{21}~{\rm cm^{-2}}$ the asymptotic peak strength decreases
as $\tau_1^{-3/2}e^{-\tau_1}$.

In the limit $y_{\rm crit}\ll -\ln\left<\tau_2\right>$ we can use the approximation
\begin{equation}
	f(\tau)\approx \sqrt{\pi} - \sqrt{\tau}
\end{equation}
which provides:
%
\begin{eqnarray}
	\label{eq_approxlow}
	\Delta\tau^{\rm eff} \approx \sqrt{\frac{2}{\pi}}\frac{e^{-\sigma^2/8}}{\sigma}&\Big\{&
		\sqrt{\left<\tau_2\right>}\left[\sqrt{\pi}(1-(\tau_1-\tau_2))-2\sqrt{\tau_2}\right]\nonumber\\
		&&-\sqrt{\left<\tau_1\right>}\left(\sqrt{\pi}-2\sqrt{\tau_1}\right)\Big\}
\end{eqnarray}
This approximation provides accurate asymptotic peak strength below $[N_{\rm H}]_{\rm crit}=10^{20}~{\rm cm^{-2}}$.

\end{document}